# Small Interplanetary Magnetic Flux Rope[*]

FENG HengQiang* ZHAO GuoQing WANG JieMin

Institute of Space Physics, Luoyang Normal University, Luoyang 471934, China

**Abstract:** Small interplanetary magnetic flux ropes (SIMFRs) are often detected by space satellites in the interplanetary space near 1 AU. These ropes can be fitted by a cylindrically symmetric magnetic model. The durations of SIMFRsare usually <12 h, and the diameters of SIMFRsare <0.20 AU and show power law distribution. Most SIMFRs are observed in the typically slow solar wind (<500 km/s), and only several events are observed with high speed (>700 km/s). Some SIMFRs demonstrate abnormal heavy ion compositions, such as abnormally high He abundance, abnormally high average iron ionization, and enhanced $O^{7+}$ abundance. These SIMFRs originate from remarkablyhot coronal origins. Approximately 74.5% SIMFRs exhibit counterstreamingsuprathermal electron signatures. Given their flux rope configuration, SIMFRs are potentially more effective for substorms. SIMFRs and magnetic clouds havemany similar observational properties but also show some different observations.These similar properties may indicate that SIMFRs are the interplanetary counterparts of small coronal mass ejections. Some direct bodies of evidence have confirmed that several SIMFRs areinterplanetary counterparts of CMEs. However, their different properties may imply that some SIMFRs haveinterplanetary origins. Therefore, one of the main aims of future research on SIMFRs is to determine whether SIMFRs originate from two different sources, that is, some events are formed in the solar coronal atmosphere, whereas others originate from the interplanetary space. Finally, in this study, we offer some prospects that shouldbe addressed in the future.

**Key Word: magnetic flux ropes, magnetic cloud, coronal mass ejection, heliospheric current sheet**

## 1 Introduction

Early researchers speculated that rapid plasma flows may exist between the sun and the earth, and such plasma flows can cause geomagnetic storms[1].These plasma flows have been later called magnetized plasma clouds, drive gas, nascent streams, magnetic tongues, and ejects[2-6].In the 1960s and 1970s,numerous observations showed that the magnetic fields within these ejects are enhanced and disordered [7]. It was soon realized that these ejections wereinterplanetary counterparts of coronal mass ejections (CMEs)[8], which are a kindof the largest eruptions on the sun.Currently, these ejections are commonly referred to as interplanetary CMEs (ICMEs), which are the main cause of aperiodic geomagnetic storms [9].By the early eighties of the last century, manyobservational characteristics of ICMEshave been recognized, such as unusually low proton

E-mail: fenghq9921@163.com

temperatures [8], counterstreaming suprathermal electrons (CSEs) [10], enhanced plasma He abundances [8,11], and enhancements in high charge states of heavy ions[12]and so on.Burlaga et al. [13] introduced the magnetic cloud (MC) to describe a special class of ICMEs with the following properties: enhanced magnetic field magnitude, smooth rotation of magnetic field direction,and decreased proton temperature.The first reported MC was observed by four satellites (IMP-8, Helios A, Helios B, and Voyager 2) on 1January 1978.According to each satellite measurements of the magnetic field, the magnetic field configurations of MCs can be described using a large cylindrically symmetric rope. Burlaga et al.[14]estimated the local axial direction and diameter ofthe flux rope and presented a schematic of this event in interplanetary space (Figure 1). The dashed part of Figure 1 indicates that the both ends of the MC may be disconnected from the sun's magnetic field lines; however, CSE observationshave revealed that the largemagnetic flux rope (MFR) structures of MCs usually remain connected with the sun's magnetic field linesat both ends near 1 AU [15].Approximately 40% of ICMEs are MCs, and the percentage varies with the solar cycle[16].MCs are often observed behind a shock and always induce major geomagnetic storms [17,18].

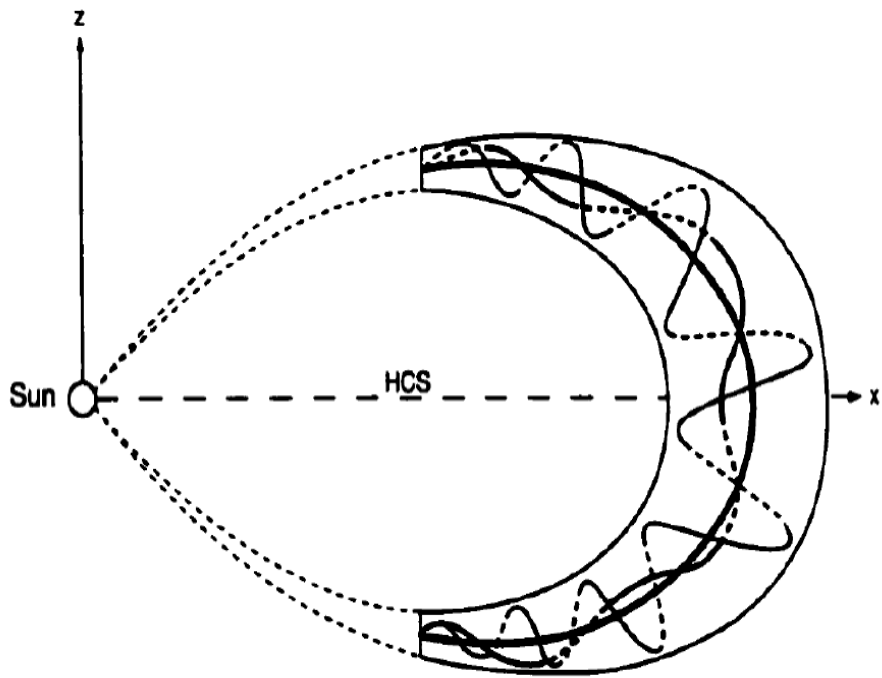

Figure 1 Schematic of the magnetic flux rope configuration and its possible relationship with the sun [14].

MCs, as large-scale interplanetary MFR (IMFR), have a spatial distribution that ranges from 0.20–0.40 AU at approximately 1 AU [19].Small IMFRs (SIMFRs) are often detected bysatellites near the earth.Compared with MCs, SIMFRsare defined empirically by Feng et al. [20] as demonstrating the following characteristics: (1) approximately flux rope magnetic field configurations and (2) diameters of <0.20 AU and durations of <12 h.MCs have consistently elicited attention and have been widely studied,and we possess a deep understanding of their solar origins, magnetic structures, evolution, and geomagnetic effect [18, 21-23]. Compared with those of MCs, the origins of SIMFRsare not yet fully understood [24]. SIMFRs were first reported by Moldwin et al.[25], who proposed that SIMFRs originate from the magnetic reconnection atthe heliospheric current sheet (HCS). The authors mainly argue that a so-calledintermediate-scale (i.e., several hours) MFR in the interplanetary space has not been reported.If SIMFRs and MCs originate in the solar corona,then all IMFRs (i.e., SIMFRs and MCs) should have acontinuous size distribution. Furthermore, Feng et al. [26] systematically investigated magnetic and plasma data measure by Wind during 1995 to 2001 and identified 144 IMFRs, and theyfound that diameters of the 144 IMFRs show continuous distribution. Hence, they proposed that all IMFRsshow the same origin and are interplanetary counterparts of CMEs or small CMEs. Since then, SIMFRs have elicited increasing attention, with focus on the interplanetary observational characteristics, possible solar origins, and geo-effectiveness [27–52].This paper provides an overview of the present understanding of SIMFRs, and presentprospects for future studies.

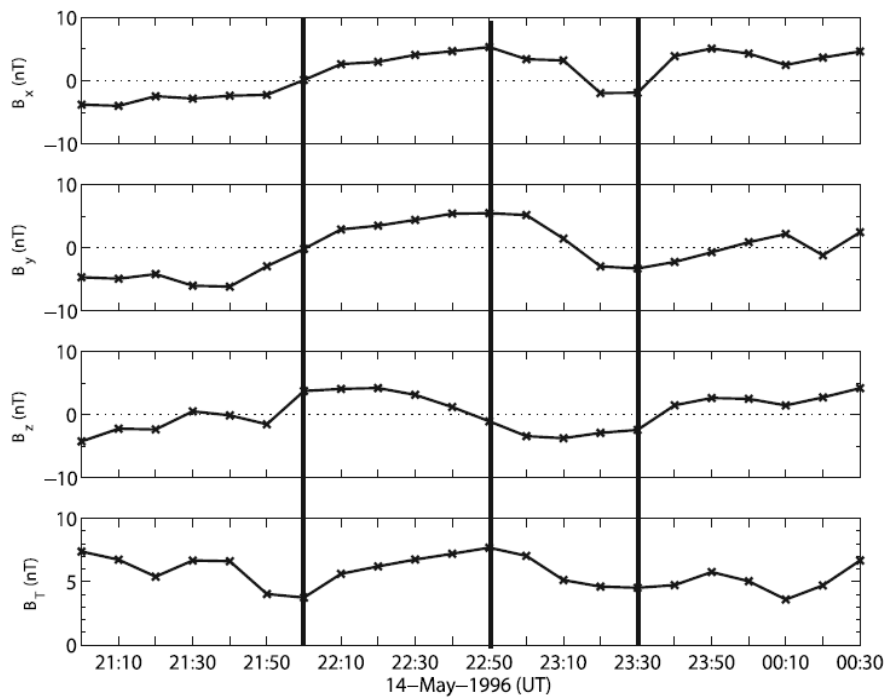

Figure 2 Magnetic field data of the SIMFR measured in the GSE coordinates on 14 May 1996.The two vertical lines at 22:00 UT and 23:30 UT denote the front and rear boundaries of the SIMFR, respectively, and the intermediate vertical line indicates the core field peak in the SIMFR [27].

**2 Interplanetary observational characteristics of SIMFRs**

All IMFRs can be fitted by using a cylindrically symmetric flux rope model [19,25]. Therefore,

the essential properties of IMFRs should be enhanced magnetic field magnitude and smooth rotation of magnetic field vector. For example, Figure 2 shows the magnetic field data of the SIMFR measured by Wind on 14 May 1996 with theduration less than 2 h [27]. From top to bottom, the panels show the x, y, and z components of the magnetic field in the GSE coordinates and total magnetic magnitude ($B_T$). In Figure 2, the essential properties of IMFRs are exhibited: the total magnetic magnitude increases slowly from 4nT at the front boundary to 8nT at 22:50 UT, the $B_x$ and $B_y$ components are enhanced at the center of the event, and the bipolar curve appears in the $B_z$ component. In the following section, we will introduce the observational and statistical characteristics of SIMFRs in details.

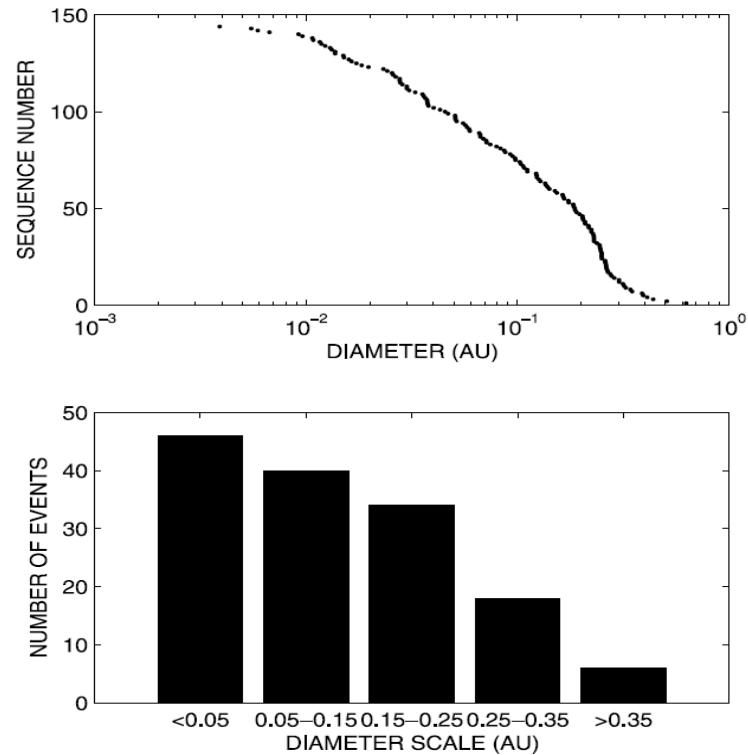

Figure 3  Diameter distribution of the 144 IMFRs observed by Wind: (top) the numberin the order of descending diameter and (bottom) the distribution of the IMFRs versus the diameter [26].

**2.1 Statistical analysis of IMFR properties**

Following the two essential properties, Feng et al. [26] selected many candidate flux rope events, fitted the candidate events by using the constant alpha and cylindrically symmetric model, and identified 144 IMFRs (i.e., SIMFRs and MCs).The diameters of these IMFRs can be estimated by using the fitting process, and the diameter distribution of the 144 IMFRs is shown in Figure 3.The numberin the order of descending diameteris shown in the upper panel, and the distribution of the IMFRs versus the diameter is shown in the lower panel.Figure 3 indicates that all IMFRs show continuous size distributions, and the occurrence rates of IMFRs decrease as the diameters increase. The physical properties of IMFRs, such as proton temperature, proton density, speed, and magnetic field strength,show no obvious change with increasing duration. Thus, all IMFRs have been suggested to be interplanetary counterparts of CMEs. Cartwright and Moldwin[27] then selected 80 SIMFRsfrom the 144 IMFRs andanalyzedthe occurrence rate of events at every hour interval.Figure 4 showsthat the duration distribution of the 80 SIMFRs appears as a double-peak structure. Given the duration distribution of all IMFRs (i.e., SIMFRs and MCs) with hourly bins, SIMFRs show power law distribution, whereas MCs demonstrate a Gaussian-like distribution [27, 41].Given that their duration distributions are different, SIMFRs and MCs may havedifferent sources.Whereas MCs originate in the sun, Cartwright and Moldwin[27]suggested that SIMFRs may be formed in the HCS.However, they did not provide convincing evidence,except that radial scale sizes of many SIMFRs are similar with the estimated HCS thickness [27]. In addition, Using Wind data from 1996–2016, Hu et al. [52] identified 74,241 small-scale magnetic flux rope events. Most of these small events have a durationof <1 h, and many events are detected within ICMEs. These small events are highly different with the SIMFRs discussed in this paper, theymay be MHD intermittent turbulence; moreover, someof their observational characteristics agree with the numerical simulation results of Greco et al.[53].

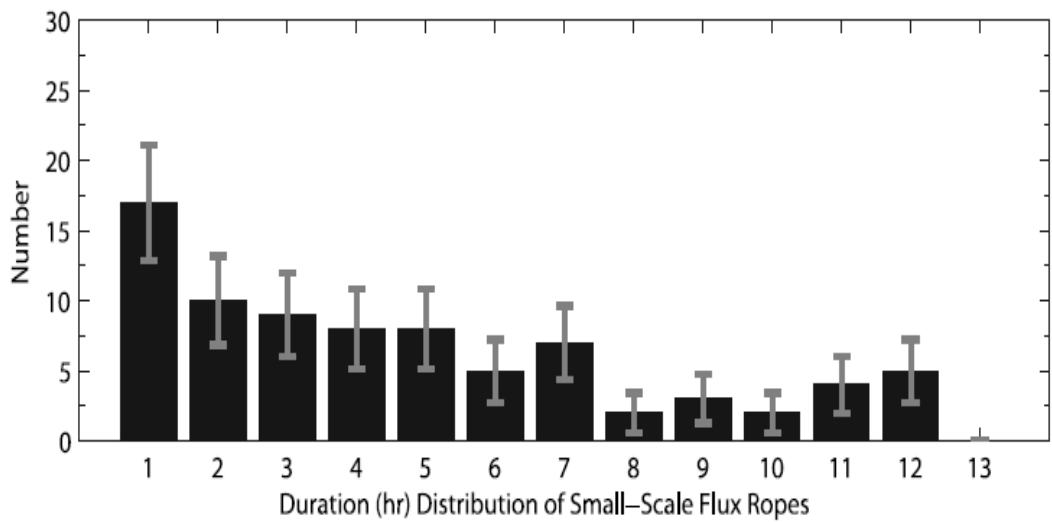

Figure 4Duration distribution of SIMFRs observed by Wind [27].

One of the main questions in SIMFR studies is the possible relation between SIMFRs and the MCs, and it is related to their source: Do SIMFRs originate in the solar corona or directly in the heliosphere?Extensive effort has been invested in systematic surveys onthe properties of SIMFRs to compare SIMFRs with MCs. The statistical results from related investigations indicate thatMCs and SIMFRs have some similar observational properties, but also haveseveral different properties. The similar properties are listed as follows. (1) SIMFRs and MCs show the approximate trends of annual occurrence rates, and the occurrence rates of MCs and SIMFRs werelow in 1999[27, 28]. However,the occurrence rates of SIMFRs and MCsdemonstrate no apparent solar cycle dependence[27, 28]. (2) Most SIMFRs and MCs are observed in the typically slow solar wind (<500 km/s), whereas onlyseveral events are observed at high speed (>700 km/s)[27, 28, 41, 54].

This finding may indicate that most IMFRs (i.e., SIMFRs and MCs)are decelerated when plowing through slow ambient solar wind because of their closed magnetic structure [54].(3) The axial distributions of SIMFRs and MCsare almost consistent and their axial longitudes (i.e., azimuthal direction in theecliptic)are scattered; however,their axial latitudes (i.e., inclination relative to the ecliptic) are predominantly within ±50degree[28, 54, 55].It may mean that both SIMFRs and MCs have similar solar source locations. (4) SIMFRs and MCs haveboundary layer structures, in which signatures of magnetic reconnection are exhibited. Their boundary layer structures may be formed by the interactions (i.e., compression and magnetic reconnection process) between the IMFRs and the interplanetary backgroundsolar wind[33, 37, 38, 56]. This result indicates that they may havesimilar propagation processes.We would like to point out thatsmall-scale flux tubes bound by current sheet-type discontinuities can be generated through MHD turbulence processes, and these small events can exhibitsimilarboundary layerstructures [53]. (5) SIMFRs and MCs demonstrate abnormal heavy ion compositions, such as abnormally high He abundance, abnormally high average iron ionization,and enhanced $O^{7+}$abundance. Theseabnormal heavy ion compositions can provide information about the temperature and composition of their solar source regions. In particular, abnormally high average iron ionizationsare clearly associated with flare heating[43]. (6) Approximately87.5% of MCsand 74.5% of SIMFRs exhibit CSE signatures [15, 44],and this finding may indicate that most IMFRsremain connected with the sun's magnetic field linesat both endsat 1 AU.These similar properties may indicate that all IMFRsthe same original mechanism and thatSIMFRs are the interplanetary counterparts of flare-associated small CMEs.

**Table 1**Differences and similarities between MCs and SIMFRs

| No. | Similarities | Differences |
|---|---|---|
| 01 | SIMFRs and MCsshow the approximate trends of annual occurrence rates | Compared with MCs, SIMFRsexhibit lower averagemagnetic magnitude and show no apparent expansion. |
| 02 | SIMFRs and MCs demonstrate the typical slow solar wind speeds | Compared with MCs, SIMFRsdemonstrate no obviouslowerproton temperature, density, and plasma beta. |
| 03 | SIMFRs and MCsshowconsistentaxial distribution | The sizes of SIMFRs have apower lawdistribution but MCs have a Gaussian-like distribution |
| 04 | SIMFRs and MCsshow boundary layer structures | |
| 05 | SIMFRs and MCsdemonstrate abnormal heavy ion compositions | |
| 06 | SIMFRs and MCs exhibit CSE signatures | |

MCs and SIMFRs havemany similar features, but they also havedifferent properties:(1) Compared with MCs, the averagemagnetic magnitude of SIMFRs is lower and SIMFRs exhibit no apparent expansion [25–27, 42, 43]. (2) Within SIMFRs or in their background solar wind, proton temperature, density, and plasma beta are approximatelyconsistent but are much lower within MCs than that in backgroundmedium [25–27]. (3) The size distribution of IMFRs appears to be bimodal, and SIMFRs exhibit power law distribution but MCs assume a Gaussian-like distribution [27, 41]. The differences and similarities between MCs and SIMFRs are listed in Table 1. As mentioned above, the similar statistical characters of SIMFRs and MCs may imply that they havethe same origin. Are the different propertysignatures of different origins? The different properties may indicate that they havedifferent formation processes, but these differences may only be due to scale effects. This situation will be discussed in detail in the next section.

**2.2 Interplanetary evolution of SIMFR properties**

Cartwright and Moldwin[31] reported some SIMFRsthat were observed by Helios 1 and Helios 2 at approximately 0.3 AU and insisted that these SIMFRs are created due to magnetic reconnection across HCS in the inner heliosphere. Therefore, regardless ofwhether SIMFRs originate at the sun or by reconnection at the HCS, they may have propagated for a certain period when they reach 1 AU. SIMFRs interact with background medium as they move in the interplanetary space. The interaction is a key issueto understand the evolution of SIMFRs [33];after a long period of propagation, some observed properties of SIMFRs can beremoved by the interactions. For example, the radial evolution of SIMFRs mainly depends on the pressure difference between theirinternal plasmas and the background solar wind [42]. Most SIMFRs exhibit signatures of high proton temperature, density, plasma beta, and depression of the magnetic field strength near their boundaries at 1 AU [33, 37, 38]. Most MCs exhibit the same signatures near their boundaries. Based on these signatures, Wei et al. [56] showed that the boundaries of MCsare not just magnetic directional discontinuitiesbut are mostly boundary layer structure. The boundary layer structures always display intensity drop and abrupt directional change in the magnetic field, relatively high proton temperature, high proton density, and high plasma beta.Wei et al.[56]concluded that these boundary layerstructuresare formed by the interactions (i.e., compression and magnetic reconnection process) between MCs and the interplanetary background solar wind. If SIMFRs came from the corona, like MCs, their early expansion will still be caused by the interactions (i.e., compression and magnetic reconnection process) between SIMFRs with the surrounding solar wind to form boundary layer structure. Thus, many SIMFRs maypossess boundary layer structures.

Feng et al.[38]surveyed plasma data and magnetic field data near the boundaries of SIMFRs and confirmed that most SIMFRs have boundary layer structures. Figure 5 shows the magnetic field data and plasma data detected by Wind during the SIMFR on the 5 March 2004 passage. In Figure 5, the two vertical dashed lines are the front and rear boundaries provided by Tian et al. [33]. Before the front boundaries, the plasma beta, proton density, and temperature are enhanced remarkably, and total magnetic magnitudeis markedly decrease. These observations indicate the presence of a boundary layer structure prior to this front boundary. The same observational features appear behind the rear boundary, that is, this SIMFR possesses a rear boundary layer. According to the reconnection exhaust criteria of Gosling et al. [57], magnetic reconnections (vertical shaded bars in Figure 5) occur within the boundary layers. The observations support that boundary layers can be formed by magnetic reconnections between SIMFRs and background magnetic fields. The boundary layer structures of SIMFRs at least indicate that SIMFRs have propagated for a long time, some observed characteristics (e.g., expansion, proton temperature, density, and plasma beta) may be modified by propagation, and their boundaries continue to evolve by interaction with background solar wind.

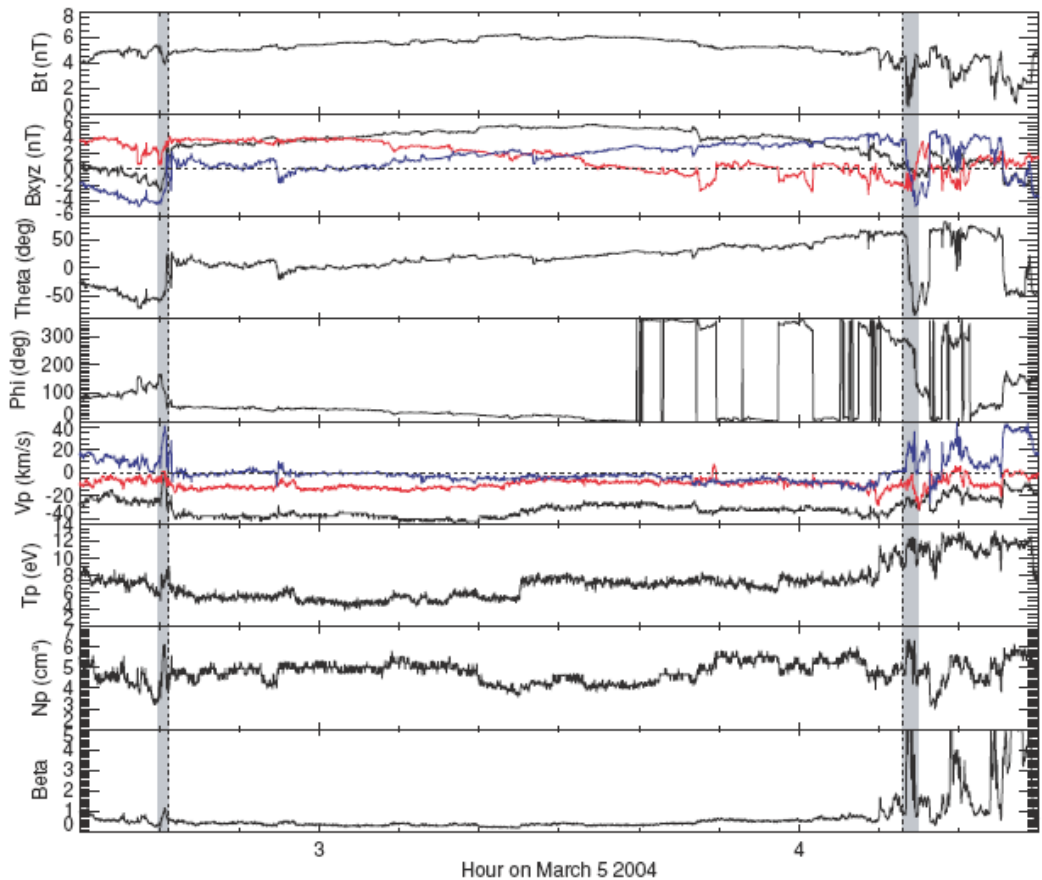

Figure 5 Magnetic field and plasma data detected by the Wind satellite during the SIMFR on the 5 March 2004 SIMFR passage. The vertical shaded bars are the observed magnetic reconnection exhaust within boundary layers[33].

SIMFRs are easily affected by background media due to their small size and lower magnetic field magnitude[42]. The difference in average magnetic magnitude between MCs and SIMFRs is only due to their different scales. Janvier et al. [42] and Feng et al. [17] found that the average magnetic magnitude of IMFRs increases with their increasing radius. The total pressure differences between SIMFRs and their surroundings are smaller than those of MCs, and the normalized expansion rate for SIMFRs is approximately half of that of MCs [42]. Therefore, as SIMFRs move away from the sun, they may have achieved pressure balance with the surrounding solar wind before they have reached 1 AU. Thus, most SIMFRs do not exhibit considerable expansion at 1AU. As mentioned above, another main difference is that low proton temperature and density are not observed within small events in most SIMFRs at 1 AU. This difference can also be explained by the lack of expansion because of quick expansion rate for MCs such that proton density and temperature in the MC drop rapidly [58]. Therefore, MCs often exhibit low temperature and density when they reach 1 AU. In summary, some different properties mentioned above can be explained by the smaller size and lower magnetic field magnitude of SIMFRs and corresponding propagation effects.

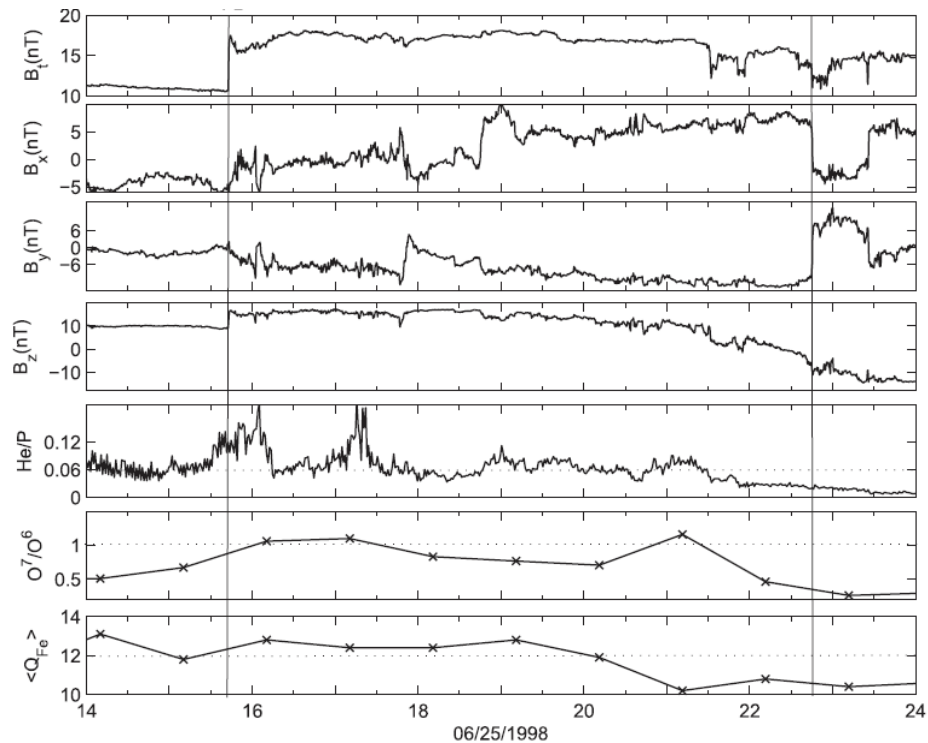

Figure 6 Magnetic field data and plasma composition detected by ACE on 25 June 1998 [43].

**2.3 Interplanetary plasma composition of SIMFRs**

Ion compositions are freezing-in near the sun if ionization and recombination timescales are larger than solar wind ion expansion time [59–60]. In the equilibrium state for solar atmospheric conditions, ion compositions are freezing-in within four solar radii [61–62]. Therefore, the ion compositions in solar wind can offer information about the temperature of their solar source regions. ICME (MCs) often demonstrate abnormal heavy ion compositions, such as abnormally high $O^{7+}/O^{6+}$ ratio and abnormally high average iron ionization $<Q_{Fe}>$ [63–67]. The average solar wind values are approximately 0.33 for $O^{7+}/O^{6+}$ ratio and approximately 10 for $<Q_{Fe}>$ [68]. If the $O^{7+}/O^{6+}$ ratio in solar wind is $\geq 1.00$ or the average iron ionization $<Q_{Fe}>$ is $\geq 12.00$, then such states are defined as abnormally high [68]. Lepri and Zurbuchen[65] found that abnormally high average iron ionizations within ICMEs are ionized by CME-associated flares. Reinard[68] obtained the same result and found that abnormally high $O^{7+}/O^{6+}$ within ICMEs are due to CME-associated flares. If SIMFRs appear withabnormally enhanced $O^{7+}/O^{6+}$ and $<Q_{Fe}>$, they may originatefrom the heating coronal regions. Figure 6 shows a SIMFR measured by ACE on 25 June1998, thefifth, sixth, and seventh panels show the He/p ratios, $O^{7+}/O^{6+}$ ratios, and average iron ionization$<Q_{Fe}>$, respectively. Figure 6 shows that the abnormally highaverage iron ionization states last for most of the SIMFR duration, and the abnormallyenhanced$O^{7+}/O^{6+}$ ratio appears in two parts. Abnormally high He abundances are often related with ICMEs, and the enhanced He abundances (He/p >0.06) are used to identify ICMEs[69-70]. Figure 6 exhibits abnormally high He/p ratio in the front half part. Feng and Wang found that SIMFRs exhibit the same high-charge-state signatures as MCs, and hot materials within SIMFRs must be heated by related flares in the corona[43]. These findings provide reliable evidence for the conjecture that SIMFRs and MCs havethe same coronal origins and thatSIMFRs are interplanetary counterparts of small CMEs [43].Huang et al. [51] discussed the formation mechanism of the twisted structures of SIMFRs by investigating their iron average charge states distributions. Compared with the five-type distributions inside large-scale MCs, only four types are confirmed among SIMFRs. Considering that multiple source regions of SIMFRs may demonstrate an effect, they also identified the possible origin of individual SIMFRs with several criteria that include signatures of$<Q_{Fe}>$ and alpha particle to proton density ratio (Nα/Np). Combined with$<Q_{Fe}>$ distributions and different source regions of SIMFRs, they implied that SIMFRs from the solar corona may assume twisted structures that are formed predominately during eruptions, and SIMFRs originating from interplanetary space may present intricate$<Q_{Fe}>$ distributions due to complex magnetic reconnection processes.

**2.4Counterstreamingsuprathermal electron signatures of SIMFRs**

Suprathermal electrons are focused along magnetic field lines called strahl, which areused to determine whether or not the interplanetary magnetic field is still connected with the sun's magnetic field lines[71]. CSEs are often observed within MCs, indicating that the flux rope structure is still connected with the sun's magnetic field linesat both ends [15]. If SIMFRs are formed in the coronal atmosphere of the sun, like MCs, the ends of most SIMFRs shouldbe still connected with the sun's magnetic field lines, and the CSE signatures shouldbe observed by satellites. Feng et al. examined 106 SIMFRsdetected by Wind to investigate their CSE signatures [44]. As an example, Figure 7exhibits the suprathermal electron distributionduring the SIMFR onthe October 12, 2004passage. The top and second panels show the 98.7and 103.3 eVsuprathermal electron distribution detected by SWE and 3DP, respectively. The second panel shows that the CSEs aredetected from 08:47 UT to11:40 UT within the SIMFR, and the interval of CSE is marked by a blue bar.Although the contrast in the top panel is not evident, the similar CSE signatures can still be observed.Feng et al. found thatnearly 75% of SIMFRs contain CSEs;thus, the CSE characteristics of SIMFRs are similar with those of MCs. This result may indicate that SIMFRs are formedinthe coronal atmosphere of the sun.

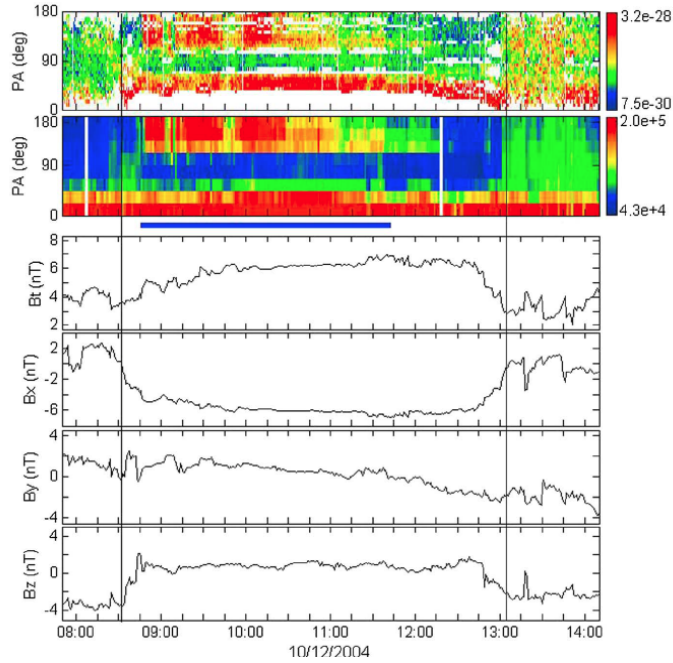

Figure 7 Suprathermal electron distribution during the SIMFR on October 12, 2004. The top and second panels show the 98.7 and 103.3 eV suprathermal electron distribution detected by SWE and 3DP, respectively. The blue bar indicates the interval of counterstreamingsuprathermal electrons [44].

Previous studies have suggested that SIMFRs may originate through magnetic reconnection in the HCS [25]. To determine the possible association between HCSs and the CSE characteristics of SIMFRs, Feng et al. [44] separated the 106 SIMFRs into two groups: the firstgroupwas measured far away from HCSs, and the second group was observed near HCSs. The statistical results show that 7 of the57 eventsand but 20 of the 49 eventsexhibit no CSEs in the first and second groups, respectively.This big difference mayindicate that thetwo groups originate from different sources. Otherwise, given that the magnetic field lines are often opened or disconnected in the vicinity of the HCS[72], the closed field lines may be openedby magnetic reconnections between SIMFRs and the magnetic field lines in the vicinity of the HCS[50].

## 3 Observations of the solar source of SIMFRs

Many similar characteristics indicate that the SIMFRs and MCs may originate in the solar coronal atmosphere. Like MCs, theycome from large-scale explosive events in the solar atmosphere,theSIMFRs may be produced bysmall-scale explosive events, such as small CMEs.In particular, some SIMFRs exhibit abnormally high ionization states as MCs, and the hot materials within SIMFRs must be heated by flares in the corona. This result indicated that at least some SIMFRsare interplanetary counterparts of small CMEs.Rouillard et al. and otherresearchers havefound corresponding direct observational evidence for some SIMFRs to come from solar explosive events [36, 73–75].

More than a decade ago, Mandrini et al. [73] reported a SIMFR with duration of 4.12 h observed by Wind on 15 May 1998. At 08:31 on11 May 1998 UT, they found an X-ray bright point near solar disc center.Asigmoid magnetic structure was then observed above the bright point on basics of the EIT 195 Å images.They provided some direct evidence for the SIMFR that linked the small eruption,namely, both the sigmoid magnetic structure and the SIMFR show the same helicity sign and magnetic field direction, and their magnetic fluxes arecomparable.The change of magnetic helicity before and after the small eruption was comparable with the total magnetic helicity of the SIMFR. All of these correlations cannot be mere coincidence, so Mandrini et al. concluded that the observed small coronal eruption resulted in the SIMFR [73].

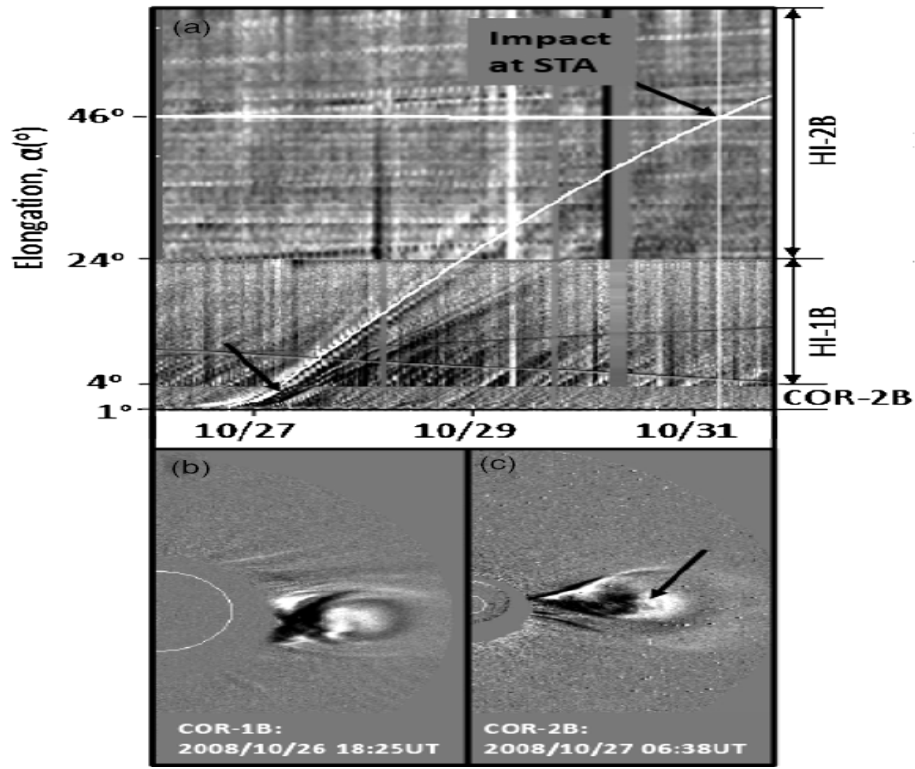

Figure 8White-light observations of the CME during 26–31October 2008 as measured by STEREO A and B [36].

Rouillard et al. [36, 74] successfully traced the origin of some SIMFRs using observations from the Sun–Earth Connection Coronal and Heliospheric Investigation (SECCHI) package [76] on board STEREO. SECCHI is composedof two heliospheric investigation cameras, coronagraphCOR-1 and COR-2, and an extreme ultraviolet imager (EUVI). The combined SECCHI cameras on two satellites enabled us to continuously track a CME from the sun to the earth. For example, Figure 8shows a bulb-shaped CME observed bySTEREO A and B on 27 October2008 (b and c) and provides calculated trajectory of the CME as a white curve.The CME emissionwas observedwith a sudden liberation of dense material observed by the EUVI instruments on STEREO A at 284 Å and STEREO B at 197 Å. The trajectory of the CME was clearly tracked by COR-1 and COR-2 on STEREO B.The latterobservations indicate that the CME propagatesapproximately along the Sun–STEREO A line. This CME was measured as a SIMFR by STEREO A in situ on 31October 2008 (Figure 9), and the duration of the related SIMFRwas approximately 5 h. Rouillard et al.[36] demonstrated several SIMFRs originating as small CMEs, which are observed by STEREO COR-1 and/or COR-2, andthese small events occur as loops or arch-like in structure.

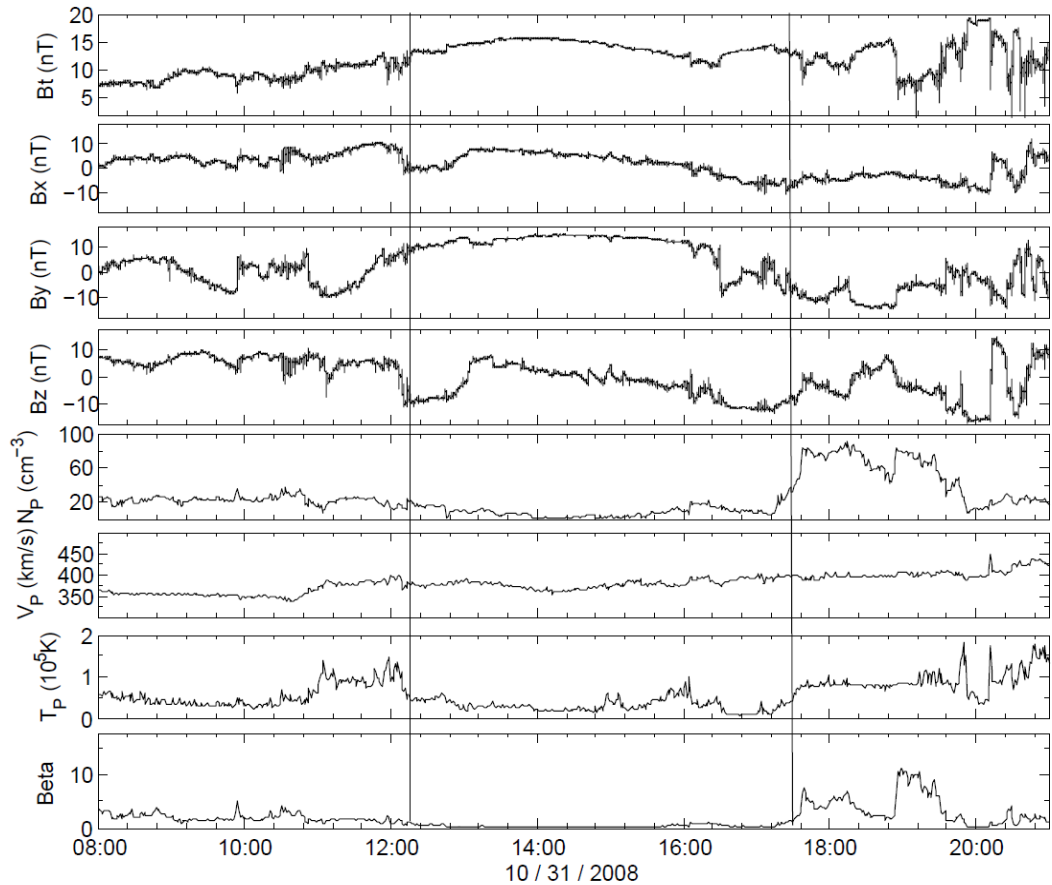

Figure 9 Magnetic field and plasma data detected by STEREO A in situ on 31 October 2008

Chi et al. [75] recently reported twosuccessive SIMFRs detected by Wind on 28 May 2011, and the durations of the two SIMFRs are approximately 6–7 h. The progenitor CMEs of the two SIMFRs were observed by SECCHI on STEREO and were launchedwithin successions of 8 h. Based on the graduated cylindrical shell (GCS) model, Chi et al. reconstructed the three-dimensional configurationsof the two CMEs, calculated theirpropagation directions, and estimated their propagation velocities and brightness. The propagation longitudes of the two CMEs are almost consistent,namely, approximately 5° west of the Sun–Earth line. Based on the near-central-meridian source locations of the two CMEs and their propagation direction, the two SIMFRs can be confirmed as interplanetary manifestations of the two CMEs. The study obtained reliable evidence to support the conjecture that some SIMFRs are interplanetary counterparts of small CMEs.

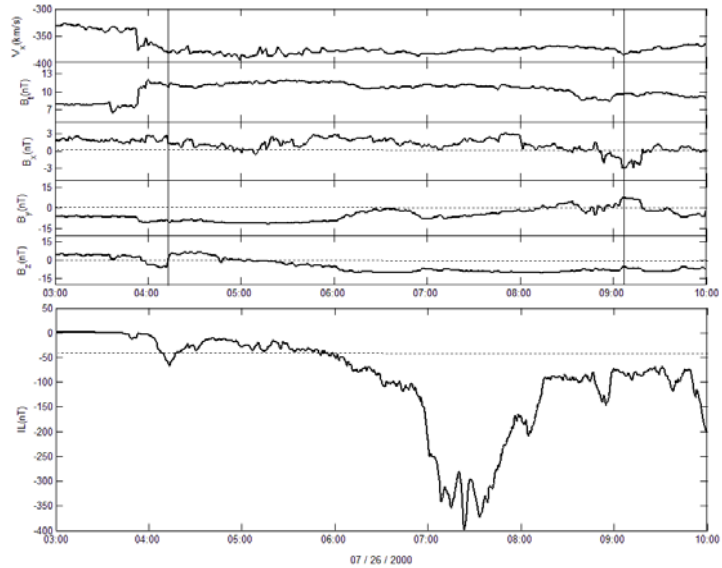

Figure 10 Sun-Earth direction velocity, magnetic field data, and IL index on July 26, 2000 [34].

**4 Geo-effectiveness of SIMFRs**

Substorms and magnetic storms are two important phenomena of geomagnetic activity [77-78].Substorms are complicated processes, and their generation mechanisms and temporal evolutions are still unclear. Some researchersidentify substorm using the IL index, which iscreated from by the IMAGE array of ground magnetometers in the northern hemisphere auroral region[79]. For each timestep, IL stands for theminimum north component of the magnetic field measured at the selected stations.A substormcan be identified when the IL index quickly drops below -80 nTand then increases slowly close to zero [80]. Southward magnetic field is a key solar wind parameter in triggering geomagnetic activity [81]. The duration and intensity of the southward magnetic field are closely related to the triggering of substorms and geomagnetic storms. The major magnetic storms can be caused by a strong and sustained southward magnetic field ($B_z < -10$ nT) [82]. In the same manner, substorms can be triggered by a short sustained southward southward magnetic field ($B_z < -3$ nT) [83]. As large-scale IMFR, MCs can provide strong and sustained southward magnetic field and are therefore effective for major geomagnetic storms [23].Similarly, SIMFRs shouldbe effective for substorms. Feng et al. [34] investigated 26 SIMFRs and found that 18 of them cause magnetosphericsubstorms.Figure 10 shows the sun–earth direction velocity, magnetic field data, andIL index history during the SIMFR eventon 26 July 2000. The Wind spacecraft detectedthe SIMFR at X =−16.7 RE. Figur e 10 shows that the IL index decreases slowly at approximately 05:20 UT and then begins to decrease rapidly at approximately 06:52 UT. A directional discontinuity (DD) is clearly measured at approximately 06:59 UT, where the Bz components turn northward. The southward magnetic fieldlasts for approximately 2 h before the DD,indicating that the substorm may be induced by the change in magnetic field orientation (northward turning)within the SIMFR.Among the 18 events, 14 substormsare triggered by the changes in magnetic field orientation (northward turning), whereas the other 4 substormsare induced by sudden enhanceddynamic pressure within SIMFRs.

Zhang et al. [39] selected 16 SIMFRs toinvestigate theirgeoeffectivenessduring solar minimum (2007–2008). The 16 SIMFRswere detected by multiple spacecraft in the near-Earth upstream ormagnetosheath, and these observations indicate that all 16 events propagated from upstream solar wind into magnetosheath.Zhang et al.[39]found that 13 of the 16 events cause magnetosphericsubstorms. Basedon published databases on SIMFRs from 1995 to 2005, Zhang et al. [39]investigated the geoeffectiveness of 141 SIMFRs, and obtained similar statistical results as follows: most of ISMFRs are effective for substorms. However, all of the substormsareinduced by sustained southward magnetic field within the SIMFRs.

**5 Summary and prospects**

In the past decades, considerable progress on SIMFRs has been achieved, and we possess a comprehensive understanding of their interplanetary observational characteristics, magnetic structures, evolution, and induced geomagnetic activity. However, we still do not clearly know the origin of SIMFRs and the relationship betweenMC and SIMFRs. Although MCs and SIMFRsshow many common or similar properties, we can only safely conclude that a small proportion of the observed SIMFRs originated from the Sun and are interplanetary counterparts of small CMEs. Although some of the different properties for SIMFRs and MCs can be explained by their different sizes and field strengths, we cannot eliminate the possibility that some SIMFRs havean interplanetary origin. Therefore, the main aim of the future research on SIMFRs is to determine whether SIMFRs havetwo different sources: some events originate from the solar coronal atmosphere, and the others are formed near HCSs in the interplanetary space.

The following aspects of work must be conducted to achieve the above aim. (1) Although some statistical studies support that SIMFRs may be formed inHCSs [24], the formation mechanism of SIMFR near the HCS remains unclear [42], and no direct evidence has been reported to support the scenario. If SIMFRs are formatted by the reconnection between the opposite magnetic field on both sides of the HCS, their axial directions shouldbe orthogonal to the local Parker spiral. However, only several cases will be oriented as expected [42]. If SIMFRs are formed in the HCS, their polarity (i.e., the temporal ordering of the north/south component, NS or SN) must change when the global solar dipole changes sign. However, the polarities of SIMFRs exhibit no changes after the global magnetic field reversal of the sun [42]. Therefore, we must employ numerical simulation to study the formation mechanism of SIMFR near the HCS and search for observational supporting evidence. (2) Some SIMFRs show abnormally high ionization states, and the related hot materials shouldbe heated by flares in the corona. On the other hand, several SIMFRs have been confirmed as interplanetary manifestations of CMEs;however, these CMEs are not associated with prominenceeruptions or flare activities. To search for evidence of solar origin, we can find flare-associated small CMEs as the origin of some SIMFRs. (3)Recently, Wang et al. [84] investigated 76 MCs observed by ACE during 1998–2007 and found that 27 (36%) events containprominence material. The predicted small CMEs may be associated with prominence eruptions. Therefore, wecan search for cool prominence materials within SIMFRs to support the solar origin of SIMFRs.(4) In the solar coronal atmosphere, many small eruptions, such as plasma blobs, plasmoids, and X-ray bright sigmoid magnetic structures, are found[85-90]. A connection between these small eruptions and SIMFRs may exist, and a survey of their connection is needed. (5) Finally, a considerable advancement on the subject of SIMFRs is expected to be provided by Parker Solar Probe. The Parker Solar Probe spacecraft can track and/or detect SIMFRs closer to the sun.

**Acknowledgments:** FENG HengQiang issupported by grant 41674170and 41804162 from the Natural Science Foundation of China.

**References:**

1 Lindeman F A. Note on the theory of magnetic storms, Philos. Mag., 1919, 38: 669–684

2 Chapman S, Bartels J. Geomagnetism, Oxford Univ. Press, New York 1940.

3 Newton, H. W. Solar flares and magnetic storms, Mon. Not. R. Astron. Soc., 1943, 103: 244–257.

4 Morrison P. Solar origin of cosmic-ray time variations, Phys. Rev., 1956, 101: 1397–1404.

5 Cocconi G, Greisen K, Morrison P, et al. NuovoCimento Suppl. Ser. X, 1958, 8:161–168.

6 Akasofu S-I, Chapman S. Solar-Terrestrial Physics, Oxford Univ. Press, New York 1972

7 Hundhausen A J, Coronal Expansion and Solar Wind, Springer, New York 1972.

8 Gosling J T, Pizzo V, Bame S J. Anomalously low proton temperatures in the solar wind following interplanetary shock waves: Evidence for magnetic bottles? J Geophys Res, 1973, 78: 2001

9 Gosling, J.T., 1990. Coronal mass ejections and magnetic flux ropes in interplanetary space. In: Physics of Magnetic Flux Ropes , 1990, 58: 343-364.

10 Bame S J, Asbridge J R, Feldman W C, et al. Bi-directional streaming of solar wind electrons greater than 80 eV: ISEE evidence for a closed-field structure within the driver gas of an interplanetary shock, Geophys Res Lett, 1981, 8: 173–176

11 Borrini G, Gosling J T, Bame S J. Feldman, Helium abundance enhancements in the solar wind, J Geophys Res, 1982, 87: 7370

12 Fenimore E E. Solar wind flows associated with hot heavy ions, Astrophys J, 1980, 235: 245、13 Burlaga L F, Sittler E, Mariani F, et al. Magnetic loop behind an interplanetary shock: Voyager, Helios, and IMP 8 observations. J Geophys Res, 1981, 86: 6673–6684

14 Burlaga L F, Lepping R, Jones J A. Global configuration ofa magnetic cloud, in Physics of Magnetic Flux Ropes, Geophys. Monogr. Ser., 1990, 58:343–364.

15 Shodhan S, Crooker N U, Kahler S W, et al. Counterstreaming electrons in magnetic clouds. J Geophys Res, 2000, 105:27261–27268

16 Richardson I G, Cane H V. The fraction of interplanetary coronal mass ejections that are magnetic clouds: Evidence for a solar cycle variation, Geophys Res Lett, 2004, 31: L18804

17 Feng H Q, Wu D J, Chao J K, et al. Are all leading shocks driven by magnetic clouds?, J Geophys Res, 2010, 115, A04107

18 Webb D F, Cliver EW, Crooker N U, et al. Relationship of halo coronal mass ejections, magnetic clouds, and magnetic storms. J Geophys Res, 2000, 105:7491–7508

19 Lepping R P, Burlaga L F,.Jones J A. Magnetic field structure of interplanetary magnetic clouds at 1 AU, J. Geophys.Res., 1990, 95: 11957– 11965.

20 Feng H Q, Wu D J, Lin C C, et al. Interplanetary small- and intermediate-sized magnetic flux ropes during

1995–2005. J Geophys Res, 2008, 113:A12105

21 Burlaga L F. Magnetic clouds and force-free fields with constant alpha. J Geophys Res, 1988, 93:7217–7224

22 Bothmer V, Schwenn R.The structure and origin of magnetic clouds in the solar wind. Ann Geophys, 1998, 16:1–24

23 Wu C C, Lepping R P, Gopalswamy N. Relationships among magnetic clouds, CMEs and geomagnetic storms. Sol Phys, 2006, 239:449–460

24 Zheng J, Hu Q. Observational evidence for self-generation of small-scale magnetic flux ropes from intermittent solar wind turbulence. Astrophys J Lett, 2018, 852:L23

25 Moldwin M B, Ford S, Lepping R, et al. Small-scale magnetic flux ropes in the solar wind. Geophys Res Lett, 2000, 27:57–60

26 Feng H Q, Wu D J, Chao J K. Size and energy distributions of interplanetary magnetic flux ropes. J Geophys Res, 2007, 112:A02102

27 Cartwright M L, Moldwin M B. Comparison of small-scale flux rope magnetic properties to large-scale magnetic clouds: Evidence for reconnection across the HCS? J Geophys Res, 2008, 113:A09105

28 Feng H Q, Wu D J, Lin C C, et al. Interplanetary small- and intermediate-sized magnetic flux ropes during 1995–2005. J Geophys Res, 2008, 113:A12105

28 Wu D J, Feng H Q, Chao J K. Energy spectrum of interplanetary magnetic flux ropes and its connection with solar activity. AstronAstrophys 2008, 480:L9-L12

29 Feng H Q, Wu D J. Observations of a small interplanetary magnetic flux rope associated with a magnetic reconnection. Astrophys J, 2009, 705:1385–1387

30 Ruan P, Korth A, Marsch E, et al. Multiple-spacecraft study of an extended magnetic structure in the solar wind. J Geophys Res, 2009, 114:A02108

31 Cartwright M L, Moldwin M B. Heliospheric evolution of solar wind small-scale magnetic flux ropes. J Geophys Res, 2010, 115:A08102

32 Gosling J T, The W –L, Eriksson S. A Torsional Alfvén Wave Embedded Within a Small Magnetic Flux Rope in the Solar Wind. Astrophys J Lett, 2010, 719:L36-L40

33 Tian Hui, Yao Shuo, ZongQiugang, et al. Signatures of Magnetic Reconnection at Boundaries of Interplanetary Small-scale Magnetic Flux Ropes. Astrophys J, 2010, 720:454-464

34 Feng H Q, Chao J K, Lyu L H, et al. The relationship between small interplanetary magnetic flux rope and the substorm expansion phase. J Geophys Res, 2010, 115:A09108

35 Feng H Q, Wu D J, Chao J K. Comment on "Comparison of small-scale flux rope magnetic properties to large-scale magnetic clouds: Evidence for reconnection across the HCS"? by M. L. Cartwright and M. B. Moldwin. J Geophys Res, 2010, 115:A10109

36 Rouillard A P, Sheeley N R, Cooper T J, et al.The Solar Origin of Small Interplanetary Transients.Astrophys J, 2011, 734:7

37 Feng H Q, Wu D J, Wang J M, et al. Magnetic reconnection exhausts at the boundaries of small interplanetary magnetic flux ropes. AstronAstrophys 2011, 527:A67

38 Feng H Q, Wang J M, Wu D J. The evidence for the evolution of interplanetary small flux ropes: Boundary layers. Chin Sci Bull, 2012, 57:1415–1420

39 Zhang X Y, Moldwin M B, Cartwright, M. The geo-effectiveness of interplanetary small-scale magnetic flux ropes. Journal of Atmospheric and Solar-Terrestrial Physics, 2013, 95:1-14

40 Yu W, Farrugia C J, Lugaz N, et al. A statistical analysis of properties of small transients in the solar wind 2007-2009: STEREO and Wind observations. J Geophys Res, 2014, 119:689–708

41 Janvier M, Démoulin P, Dasso S. Are There Different Populations of Flux Ropes in the Solar Wind? Sol Phys, 2014, 289:2633-2652

42 Janvier M, Démoulin P, Dasso S. In situ properties of small and large flux ropes in the solar wind. J Geophys Res, 2014, 119:7088-7107

43 Feng H Q, Wang J M. Observations of Several Unusual Plasma Compositional Signatures within Small Interplanetary Magnetic Flux Ropes. Astrophys J, 2010, 809:112

44 Feng H Q, Zhao G Q, Wang J M. Counterstreaming electrons in small interplanetary magnetic flux ropes. J Geophys Res, 2015, 120:10175-10184

45 Yu W, Farrugia C J, Galvin A B, et al. Small solar wind transients at 1 AU: STEREO observations (2007-2014) and comparison with near-Earth wind results (1995-2014). J Geophys Res, 2016, 121:5005-5024

46 Zheng J, Hu Q. Observations and analysis of small-scale magnetic flux ropes in the solar wind. Journal of Physics: Conference Series, 2016, 767:012028

47 Zheng J, Hu Q, Chen Y, et al. Automated Detection of Small-scale Magnetic Flux Ropes and Their Association with Shock. Journal of Physics: Conference Series, 2017, 900:012024

48 Huang J, Liu Y, Peng J, et al. A multispacecraft study of a small flux rope entrained by rolling back magnetic field lines. J Geophys Res, 2017, 122:6927-6939

49 Wang J M, Liu Q, Zhao Y. Magnetic Disconnections at the Boundary of a Small Interplanetary Magnetic Flux Rope Associated with a Reconnection Exhaust. Sol Phys, 2018, 293:116

50 Wang J M, Feng H Q, Zhao G Q. Observations of a Small Interplanetary Magnetic Flux Rope Opening by Interchange Reconnection. Astrophys J, 2018, 853: 94

51 Huang J, Liu Y, Peng J, et al. (2018). The distributions of iron average charge states in small flux ropes in interplanetary space: Clues to their twisted structures. J Geophys Res, 2018, 123: 7167–7180.

52 Hu Q, Zheng J, Chen Y, et al. Automated Detection of Small-scale Magnetic Flux Ropes in the Solar Wind: First Results from the Wind Spacecraft Measurements. Astrophys J S, 2018, 239: 12

53 Greco A, Chuychai P, Matthaeus W H, et al. Intermittent MHD structures and classical discontinuities. Geophys Res Lett, 2008, 35, L19111

54 Bothmer V, Schwenn R.The structure and origin of magnetic clouds in solar wind. Ann Geophys, 1998, 16: 1–24

55 Lepping R P, Berdichevsky D. (2000), Interplanetary magnetic clouds: Sources, properties, modeling, and geomagnetic relationship. Recent Res Dev Geophys, 2000, 3:77–96

56 Wei F, Liu R, Fan Q, et al. Identification of the magnetic cloud boundary layers. J Geophys Res, 2003, 108: 1263

57 Gosling J T, Skoug R M, McComas D J, et al. Direct evidence for magnetic reconnection in the solar wind near 1 AU. J Geophys Res, 2005,110: A01107

58 Wang C, Du D, Richardson J D. Characteristics of the interplanetary coronal mass ejections in the heliosphere between 0.3 and 5.4 AU. J Geophys Res, 2005, 110: A10107

59 Heidrich-Meisner V, Peleikis T, Kruse, M, et al. Observations of high and low Fe charge states in individual solar wind streams with coronal-hole origin. AstronAstrophys 2016, 593:A70

60 Wang J M, Feng H Q. Observational connection between local high-temperature phenomena within magnetic clouds and the Sun. Science China Earth Sciences, 2016, 59: 1051–1056,

61 Hundhausen A J, Gilbert H E, Bame S J.The State of Ionization of Oxygen in the Solar Wind.Astrophys J, 1968, 152: L3

62 Bame S J, Asbridge J R, Feldman W C, et al. The Quiet Corona: Temperature and Temperature Gradient. Sol Phys, 1974, 35: 137-152

63 Buergi A, Geiss J. Helium and minor ions in the corona and solar wind - Dynamics and charge states. Sol Phys, 1986, 103: 137-152

64 Lepri S T, Zurbuchen T H, Fisk L A, et al. Iron charge distribution as an identifier of interplanetary coronal

mass ejections. J Geophys Res, 2001, 106: 29231–29238

65 Lepri S T, Zurbuchen T H. Iron charge state distributions as an indicator of hot ICMEs: Possible sources and temporal and spatial variations during solar maximum. J Geophys Res, 2004, 109: A01112

66 Song H Q, Chen Y, Zhang J, et al. Evidence of the Solar EUV Hot Channel as a Magnetic Flux Rope from Remote-sensing and In Situ Observations. Astrophys J Lett, 2015, 808:L15

67 Song H Q, Zhong Z, Chen Y, et al. A Statistical Study of the Average Iron Charge State Distributions inside Magnetic Clouds for Solar Cycle 23.Astrophys J S, 2016, 224:27

68 Reinard A. Comparison of interplanetary CME charge state composition with CME-associated flare magnitude. Astrophys J, 2005, 620:501–505

69 Burlaga L F, Skong R M, Smith C W, et al. Fast ejecta during the ascending phase of solar cycle 23: ACE observations, 1998–1999. J Geophys Res, 2001, 106: 20957–20978

70 Elliott H A, McComas D J, Schwadron N A, et al. An improved expected temperature formula for identifying interplanetary coronal mass ejections, J Geophys Res, 2005, 110: A04103

71 Pagel C, Crooker N U, Larson D E, et al. Understanding electron heat flux signatures in the solar wind. J Geophys Res, 2005, 110: A01103

72 Gosling J T, Skoug R M, McComas D J, et al. Magnetic disconnection from the Sun: Observations of a reconnection exhaust in the solar wind at the heliospheric current sheet. Geophys Res Lett, 2005, 32, L05105

73 Mandrini C H, Pohjolainen S, Dasso S, et al. Interplanetary flux rope ejected from a X-ray bright point: The smallest magnetic cloud source-region ever observed. AstronAstrophys, 2005, 434: 725–740

74 Rouillard A P, SavaniJ N P, Davies A, et al.A Multispacecraft Analysis of a Small-Scale Transient Entrained by Solar Wind Streams. Solar Phys, 2009, 256: 307

75 Chi Y, Zhang J, Shen C, et al. Observational Study of an Earth-affecting Problematic ICME from STEREO. Astrophys J, 2018, 863: 108

76 Howard R A, Moses J D, Vourlidas A, et al. Sun Earth Connection Coronal and Heliospheric Investigation (SECCHI), Space Sci. Rev, 2008, 136: 67-115

77 Akasofu S I.The development of the auroralsubstorm. Planet Space Sci, 1964, 12: 273–282

78 McPherron R L. Magnetosphericsubstorms, Rev Geophys, 1979, 17: 657–681

79 Kauristie K, Pulkkinen T I, Pellinen R J, et al. What can we tell about auroralelectrojet activity from a single meridional magnetometer chain?, Ann Geophys, 1996, 14: 1177.

80 Tanskanen E I. A comprehensive high‐throughput analysis of substorms observed by IMAGE magnetometer network: Years 1993–2003 examined, J Geophys Res, 2009, 114: A05204

81 Tavares M, Christensen F, Moretto T, et al. Semiannual variation of geomagnetic activity in the Greenland magnetometer chain. Phys Chemistry of the Earth, 1997, 22: 685.

82 Gonzalez W D, Tsurutani B T. Criteria of interplanetary parameters causing intense magnetic storms (Dst< −100 nT), Planet Space Sci, 1987, 35:1101–1109

83 Kamide Y, Perreault P D, Akasofu S I, et al. Dependence of substorm occurrence probability on interplanetary magnetic field and on size of auroral oval. J Geophys Res, 1977, 82: 5521–5528.

84 Wang J M, Feng H Q, Zhao G Q. Cold prominence materials detected within magnetic clouds during 1998–2007. AstronAstrophys 2018, 616:A41

85 Chen Y, Li X, Song H Q, et al. Intrinsic instability of coronal ctreamers. Astrophys J, 2009, 691: 1936–1942

86 Song H Q, Kong X L, Chen Y, et al.A Statistical study on the morphology of rays and dynamics of blobs in the Wake of coronal mass ejections. Sol Phys, 2011, 276: 261-276

87 Song H Q, Chen Y, Liu K, et al. Quasi-periodic releases of streamer blobs and velocity variability of the slow solar wind near the Sun. Sol Phys, 2009, 258: 129–140

88 Sterling A C, Hudson H S. YOHKOH SXT observations of X-ray "dimming" associated with a halo coronal mass ejection. Astrophys J, 1997, 491: L55–L58

89 Ohyama M, Shibata K. X-ray plasma ejection associated with an impulsive flare on 1992 October 5: Physical conditions of X-ray plasma ejection. Astrophys J, 1998, 499: 934–944

90 Shibata K, Masuda S, Shimojo M, et al. Hot-plasma ejections associated with compact-loop solar flares. Astrophys J, 1995, 451: L83–L85